\documentclass[twocolumn]{svjour3}
\usepackage[utf8]{inputenc}
\usepackage[T1]{fontenc}
\usepackage{amssymb,amsmath}
\usepackage{etoolbox}
\usepackage{xspace}
\usepackage{subcaption}
\usepackage{graphicx}
\usepackage{color}
\usepackage[version=4]{mhchem}
\usepackage{relsize}
\usepackage{microtype}
\usepackage{siunitx}
\usepackage{soul}

\usepackage{hyperref}
\usepackage[
    backend=biber,
    sorting=none,
    style=numeric-comp,
    natbib=true,
    hyperref=true,
    doi=false,
    isbn=false,
    url=false,
]{biblatex}

\newcommand{\sublabel}[2]{%
    \setbox1=\hbox{#2}  
    \leavevmode\rlap{\usebox1}  
    \rlap{\hspace*{10pt}\raisebox{\dimexpr\ht1}{#1}}  
    \phantom{\usebox1}  
}

\bibliography{OFAST}

\DeclareCiteCommand{\citenum}
    {}
    {\printfield{labelnumber}}
    {}
    {}

\DeclareCiteCommand{\cite}[\mkbibsuperscript]
  {\bibopenbracket
   \usebibmacro{cite:init}%
   
   \iffieldundef{prenote}
     {}
     {\BibliographyWarning{Ignoring prenote argument}}%
   \iffieldundef{postnote}
     {}
     {\BibliographyWarning{Ignoring postnote argument}}}
  {\usebibmacro{citeindex}%
   \usebibmacro{cite:comp}}
  {}
  {\usebibmacro{cite:dump}%
   \bibclosebracket}

\def\d{\text{d}}
\def\Cudhbc{Cu(dhbc)$_\text{2}$(4,4$^\prime$-bpy)\xspace}
\def\RPMZn{RPM3-Zn\xspace}

\begin{document}


\title{On the use of the IAST method for gas separation studies in porous materials with gate-opening behavior}

\author{Guillaume Fraux \and Anne Boutin \and Alain H. Fuchs \and François--Xavier Coudert}

\institute{Guillaume Fraux
    \and Alain H. Fuchs
    \and François--Xavier Coudert \at
    Chimie ParisTech, PSL Research University, CNRS, Institut de recherche de Chimie Paris, 75005 Paris, France
    \email{fx.coudert@chimie-paristech.fr}
    \and
    Anne Boutin \at
    École Normale Supérieure, PSL Research University, Département de Chimie, Sorbonne Universités -- UPMC Univ. Paris 06, CNRS UMR 8640 PASTEUR, 24, rue Lhomond, 75005 Paris, France
}

\date{}

\maketitle

\begin{abstract}
Highly flexible nanoporous materials, exhibiting for instance gate opening or
breathing behavior, are often presented as candidates for separation processes
due to their supposed high adsorption selectivity. But this view, based on
``classical'' considerations of rigid materials and the use of the Ideal
Adsorbed Solution Theory (IAST), does not necessarily hold in the presence of
framework deformations. Here, we revisit some results from the published
literature and show how proper inclusion of framework flexibility in the osmotic
thermodynamic ensemble drastically changes the conclusions, in contrast to what
intuition and standard IAST would yield. In all cases, the IAST method does not
reproduce the gate-opening behavior in the adsorption of mixtures, and may
overestimates the selectivity by up to two orders of magnitude.
\end{abstract}

\section{Introduction}

Gas separation is an important step in multiple industrial processes, from
separation of hydrocarbons in oil chemistry to \ce{CO2} separation and storage
or oxygen extraction in the air. The two main methods used for gas separation
are cryogenic distillation, mainly used for air separation, and differential
adsorption. Adsorption-based processes for gas separation, which rely on
microporous materials as an adsorber bed, are very versatile because of the
large choice of materials available --- and the possibility to tune them for a
specific system. Among the porous materials used commercially, one can list
inorganic materials (such as zeolites and silica gels), carbon-based compounds
(e.g., activated carbon), and hybrid organic--inorganic materials, including the
topical family of metal--organic frameworks (MOFs) or porous coordination
polymers (PCPs).

Experimental characterization of the co-adsorption of a mixture of gases inside
a porous adsorbent is typically done through multi-component gas adsorption
studies. This problem is inherently high-dimensional, e.g., for a ternary
mixture there are four variables to vary (temperature, total pressure, and two
independent variables for the mixture composition). Because such experimental
studies of coadsorption equilibrium thermodynamics are typically long and
expensive, there has been a great expense of literature devoted to theoretical
models for the prediction of mixture co-adsorption based on single-component
adsorption data. The most commonly used method in the field is the Ideal
Adsorbed Solution Theory (IAST),\cite{myers1965} which is relatively simple to
implement and robust, and allows the prediction of multi-component adsorption
behavior from individual single-component isotherms. In particular, it is used
to predict the potential selectivity of materials based on simple measurements
of pure component isotherms.

A novel development in the area of nanoporous materials is the increasing number
of flexible materials,\cite{coudert2015} or \emph{soft porous
crystals},\cite{horike2009} that can exhibit significant changes in
structure upon adsorption of guest molecules. Those materials, which undergo
large-scale reversible structural transitions impacting their total volume or
internal pore volume, appear to be particular common among metal--organic
frameworks based on relatively weaker bonds (coordination bonds, $\pi$--$\pi$
stacking, hydrogen bonds, or some covalent bonds) compared to inorganic dense
nanoporous materials (such as zeolites). In particular, some of these
materials show transitions between an ``open'' phase with large pore volume, and
a ``condensed'' or ``narrow pore'' phase with smaller pore volume --- or, in
some cases, no microporosity at all. Such transitions, known as \emph{gate
opening}\cite{kitaura2003, Tanaka2008, li2001}
or \emph{breathing}\cite{Serre2002, bourrelly2005} depending on the order in which the phases occur
upon adsorption, can lead to stepped adsorption isotherms. In the recent
literature, many authors have relied on IAST predictions to predict that several
such flexible MOFs would present very good selectivity for gas separation. For
example \citeauthor{nijem2012} \cite{nijem2012} reported that \emph{``[their]
work unveils unexpected hydrocarbon selectivity in a flexible metal--organic
framework (MOF), based on differences in their gate opening pressure.''} In some
cases of the published literature, the authors explicitly used IAST to derive
such predictions on flexible materials.\cite{banerjee2015, mukherjee2015,
foo2016, li2016} In other cases, IAST was not used explicitly, but the
assumptions made for the behavior of mixtures stem from the ``classical''
understanding of selectivity rules in rigid materials, and would not necessarily
be valid in flexible materials.\cite{gucuyener2010, inubushi2010, nijem2012,
sanda2013, joarder2014, mukherjee2014}

In this paper, we look at the hypotheses of the IAST method and show why they
are not fulfilled when adsorption takes place in flexible nanoporous materials.
We summarize an alternative method, the Osmotic Framework Adsorbed Solution
Theory (OFAST),\cite{coudert2009-1} that can be used when structural transitions
occur upon adsorption. We then compare the results of IAST and OFAST on two sets
of adsorption data from the published literature on gate-opening materials, and
show that the IAST method gives unrealistic results: it does not reproduce the
gate-opening behavior upon mixture adsorption, and overestimates the selectivity
by up to two orders of magnitude.

\section{Predicting multi-components adsorption}

\subsection{Ideal Adsorbed Solution Theory}

The Ideal Adsorbed Solution Theory (IAST) starts by assuming that for a given
adsorbent and at fixed temperature $T$, the pure-component isotherms $n_i(P)$
for each gas $i$ of interest is known. Then, given a mixture of ideal gases
adsorbing at total pressure $P$ in an host framework and the composition of the
gas phases $(y_i)$ --- such that the partial pressures are $P_i = y_i P$ --- the
goal of the method is to predict the total adsorbed quantity $n_\text{tot}$ and
the molar fractions $(x_i)$ in the adsorbed phase.

In order to do so, \citeauthor{myers1965}\cite{myers1965} introduced for each
mixture component a quantity homogeneous to a pressure, $P_i^*$. The IAST method
links this pressure to the compositions of the gas and adsorbed phases with two
equations for each component:
\begin{equation} P y_i = P_i^* x_i ;\label{eq:spreading}\end{equation}
for all $i$ and $j$,
\begin{equation}
    \int_0^{P_i^*} \frac{n_i(p)}{p} \d p = \int_0^{P_j^*} \frac{n_j(p)}{p} \d p \label{eq:chem-pot}.
\end{equation}
Equation~\eqref{eq:spreading} defines the link between $P_i^*$
the total pressure $P$, the gas phase molar fraction $y_i$ and the adsorbed
phase molar fraction $x_i$. Equation~\eqref{eq:chem-pot} is an expression of the
equality of chemical potentials at thermodynamic equilibrium.

In the simpler case of two-component gas mixture (B, C), these two equations and
the conservation of matter, can be rewritten to a set of four equations:
\begin{equation} P y_B = P_B^* x_B \end{equation}
\begin{equation} x_B = \frac{P_C^* - P}{P_C^* - P_B^*} \end{equation}
\begin{equation}
    \frac1{n_\text{tot}} = \frac{x_B}{n_B(P_B^*)} + \frac{1 - x_B}{n_C(P_C^*)}
\end{equation}
\begin{equation} \label{eq:iast2}
    \int_0^{P_B^*} \frac{n_B(p)}{p} \d p = \int_0^{P_C^*} \frac{n_C(p)}{p} \d p
\end{equation}
Solving these equations for $P_B^*$ and $P_C^*$ will give all the information on
the system composition. It can be done with either numerical integration of the
isotherms, or by fitting the isotherms to a model, and then integrating the
model analytically.

The IAST model for the prediction of coadsorption of mixtures in
nanoporous materials is no panacea, and more involved theories have been
developped for cases where ideality cannot be assumed: nonideal adsorbed
solution models\cite{Yang_book, NIAST_Quirke} the vacancy solution theory
(VST),\cite{VST} etc. However, IAST has been extensively studied and both its
areas of validity and its weaknesses have been well assessed. In particular, it
is known to be fairly reliable for adsorption of small gas molecules, or
mixtures of apolar fluids of a similar chemical nature (such as mixtures of
hydrocarbons). However, one limitation is that if there are big differences in
the sorption capacity, extrapolations to high pressures are necessary and thus,
the resulting mixture behavior predicted can be far off.

\subsection{IAST and flexible frameworks}

The original derivation of the IAST equations\cite{myers1965} highlights three
hypotheses on the co-adsorption process, on which the model is built:
\begin{description}
    \item[(h1)] The adsorbing framework is inert from a thermodynamic point of view;
    \item[(h2)] The adsorbing framework specific area is constant with respect to
                temperature and the same for all adsorbed species;
    \item[(h3)] The Gibbs definition of adsorption applies.
\end{description}

While the meaning of the last assumption \textbf{(h3)} has been diversely
interpreted by different authors, \citeauthor{myers1965} originaly meant it to
qualify the method by which the adsorption isotherms are measured. There is,
however, consensus on the fact that absolute adsorption should be used in IAST
calculations --- as opposed to excess or net adsorption.\cite{myers2014,
brandani2016} This assumption thus applies equally to both rigid and flexible
adsorbents. However, the first two hypothesis are not valid for flexible
nanoporous materials. \textbf{(h2)} is clearly invalid, as modifications in both
the host's volume and internal structure lead to variations of pore size and
specific area upon structural transitions. We note here, in passing, that
\textbf{(h2)} should already be ruled out for systems of pore size close to the
adsorbate diameter, as well as gas mixtures of widely different size or shape.
It should, for example, not apply to molecular sieves systems, yet those can
often be described reasonably well by IAST in practice. Finally, \textbf{(h1)}
is violated by all the systems that feature adsorption-induced deformation, and
in particular by systems presenting a gate-opening or a breathing behavior. As a
conclusion, IAST has no theoretical foundation for those systems and should not
be used for co-adsorption prediction in flexible frameworks.

\begin{figure}[ht]
    \centering
    %
    %
    %
    %
    %
    %
    %
    %
    \input{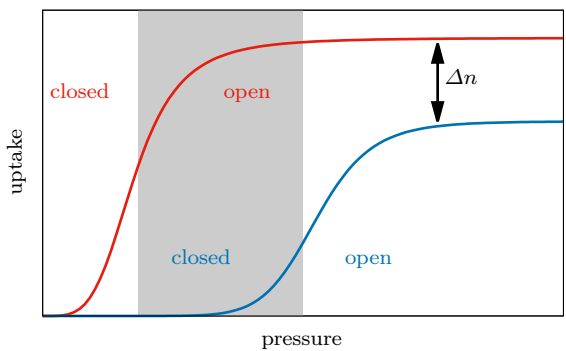}
    \caption{Typical single-component isotherms for adsorption of two gases (red
    and blue) in a material with gate opening. The gate opening pressure is not
    the same for the two adsorbates, creating a pressure range with a high
    difference in the adsorption capacity for single-components isotherms (gray
    zone in the figure). Contrary to intuition, selectivity will not necessarily
    by high in this pressure range, but will depend on difference in saturation
    uptake $\Delta n$.}
    \label{fig:open-close-selectivity}
\end{figure}

Aside from the mathematical treatment and thermodynamic hypotheses, we can show
in an qualitative way why it is not possible, in flexible host frameworks, to
use the single-component isotherm directly to predict multi-components
adsorption. We address here a common misconception, due to an invalid graphical
interpretation of the isotherms. Figure~\ref{fig:open-close-selectivity} depicts
the equilibrium adsorption isotherms for two different guests in a material
presenting a gate-opening behavior. The gate opening is an adsorption-induced
structural transition from a nonporous to a porous phase of the host, leading to
a step in the single-component adsorption isotherm. Gate opening occurs at two
different pressures for the two adsorbates, due to the specific host--guest
interactions of the two gases (characterized notably by the enthalpy of
adsorption and saturation uptake). In the pressure range in-between the
transition pressures (in gray in Figure~\ref{fig:open-close-selectivity}), the
uptake of one species is close to 0 --- \emph{in the single-component isotherm} ---
and the uptake of the other species is close to its maximum value. If these
isotherms were encountered for a rigid host material, the selectivity would be
extremely high in this range, with one guest adsorbing but not the other.

Yet, the step in the isotherms here is not simply linked to host--guest
interactions but indeed due to a change in the host structure. In particular,
upon adsorption of a gas mixture in this gate-opening framework, a phase
transition will occur at a given pressure. Before this transition, the
structure will be contracted and show no (or little) adsorption for either
guest, and thus no usable selectivity. After the transition, \emph{both species}
will adsorb into the open pore framework. The selectivity is then governed ---
at least qualitatively --- by the respective saturation uptakes of the two
fluids ($\Delta n$ in the figure). While the difference in adsorbed quantities
in the intermediate pressure range visually suggests great selectivity, it is
not possible for one component to adsorb inside the close phase framework while
at the same time the other component adsorb inside the open phase of the
framework. The framework is either in one phase or in the other, at any given
time.

\begin{figure}[ht]
    \centering
    %
    %
    %
    %
    %
    \input{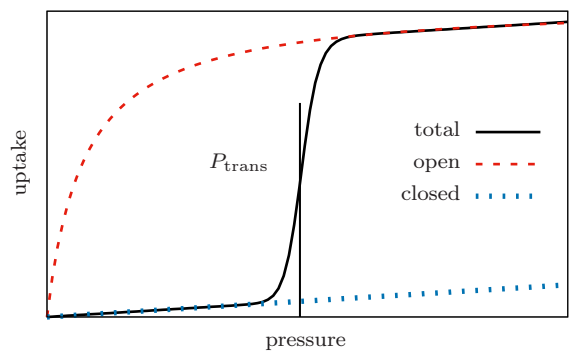}
    \caption{Generation of the total isotherm in gate-opening materials by the
    combination of two single-phase isotherms: an \emph{open} pores isotherm,
    and an \emph{closed} pores isotherm. The transition between the two host
    phases occurs at $P_\text{trans}$.}
    \label{fig:open-close-isotherms}
\end{figure}

The whole issue with using single-component isotherms to predict multi-component
adsorption in frameworks with phase transition boils down to the origin of the
stepped isotherms. The single-component isotherm (represented in
Figure~\ref{fig:open-close-isotherms}) is a combination of two isotherms: one in
the first phase (the contracted pore phase), and one in the second phase (the
open pore phase). Both phases --- and the thermodynamic equilibrium between them ---
need to be taken in account to predict the multi-component adsorption.

\subsection{The OFAST theory}

The thermodynamic ensemble suited for the study of adsorption in flexible
materials is the so-called ``osmotic ensemble'', first introduced in
\citeyear{mehta1994}\cite{mehta1994} for the study of fluid mixtures, and
adapted to multi-components phase equilibrium in
\citeyear{escobedo1998}\cite{escobedo1998}. The thermodynamic potential $\Omega$
associated with this ensemble is a function of the mechanical pressure $P$, the
temperature $T$, the number of atoms in a given host phase $\alpha$ and the
adsorbed species chemical potentials $\mu_i$:
\begin{equation}
    \Omega(T, P, \mu_i) = F_\alpha + P V_\alpha - \sum_i \mu_i N_i,
\end{equation}
where $F_\alpha$ is the Helmholtz free energy of the empty host in phase
$\alpha$, $V_\alpha$ the volume of the host in this phase, and $N_i$ the molar
uptake of guest $i$. This expression can be reworked and expressed as a function
not of chemical potentials, but of fluid pressure (taken equal to mechanical
pressure $P$) and adsorption isotherms:\cite{coudert2008}
\begin{equation} \label{eq:osmotic}
    \Omega(T, P, \mu_i) = F_\alpha + P V_\alpha - \sum_i \int_0^P n_i(T, p) V^m_i(T, p) \ \d p
\end{equation}
Here, $n_i(T,P)$ are the coadsorption isotherms for each component and
$V^m_i(T,P)$ the molar volume for the species $i$ in the bulk phase. If we
supposed that the gases are ideal, the molar volume is given by $RT/P$, with $R$
the ideal gas constant.

We have shown above that IAST cannot be used for the study of co-adsorption in
frameworks with adsorption-induced phases transition, because the framework is
not inert during adsorption. However, the IAST assumptions are still valid for
each individual phase of the host matrix, if they are considered in the absence
of a transition. As a consequence, it means that the IAST model can be used, for
each possible host phase $\alpha$, to calculate the co-adsorption isotherms
$n_{\alpha,i}(P,T)$ in this given phase. Then, the thermodynamic potential of
each phase $\Omega_{\alpha}$ can be calculated from these isotherms through
Equation~\ref{eq:osmotic}, allowing to predict which phase is the more stable at
a given gas phase pressure and composition --- and where the structural
transition(s) occur. This method, extending the IAST theory in the osmotic
ensemble to account for host flexibility, is called Osmotic Framework Adsorbed
Solution Theory (OFAST).\cite{coudert2009-1, coudert2010} Although the
amount of published data from direct experimental measurements of coadsorption
of gas mixtures in flexible MOFs is very limited, the OFAST method has been well
validated in the past against experimental data.\cite{ortiz2012, hoffmann2011,
zang2011}

In practice, the use of OFAST follows the following steps. First, the host
phases of interest are identified and the single-component adsorption isotherms
$n_{\alpha,i}(T, p)$ for these are obtained: this can be achieved from a fit of
experimental isotherms (see figure~\ref{fig:open-close-isotherms}) or from
molecular simulation.

Secondly, the relative free energies of the host phases (which reduces to a
single $\Delta F_\text{host}$ in our case of two host phases) can be computed
from equation~\eqref{eq:osmotic} and the experimental single-component stepped
isotherm. For example, with two phases $\alpha$ and $\beta$, and
considering ideal gas, we can express equation~\eqref{eq:osmotic} for each
phase:
\begin{equation}
    \Omega_\alpha(T, P, \mu_i) = F_\alpha + P V_\alpha - RT \sum_i \int_0^P \frac{n_{\alpha, i}(p)}{p} \ \d p
\end{equation}
\begin{equation}
    \Omega_\beta(T, P, \mu_i) = F_\beta + P V_\beta - RT \sum_i \int_0^P \frac{n_{\beta, i}(p)}{p} \ \d p
\end{equation}
At the transition ($P=P_\text{trans}$ in figure~\ref{fig:open-close-isotherms},
which is typically known experimentally)
the two thermodynamic potentials will be equal, which gives us a way to
evaluate the free energy difference between the phases:
\begin{equation} \label{eq:delta-f}
    \Delta F_\text{host} = RT \sum_i \int_0^{P_\text{trans}} \frac{\Delta n_i(T, p)}{p} \d p - P_\text{trans} \Delta V_\text{host}
\end{equation}

Then, for all values of thermodynamic parameters of interest (pressure and gas
mixture composition) the osmotic potential of the host phases is computed,
enabling the identification of the most stable phase: the phase with the
lowest osmotic potential is the most stable at this pressure and composition.
The pressure at which the osmotic potential in both phases are equal is the
phase transition pressure for a given composition.

Finally, we can compute adsorption properties (guest uptake and
selectivity) using IAST in this most stable phase.

\section{Results and discussion}

\begin{figure}[h]
    \centering
    \begin{subfigure}[c]{.45\textwidth}
        \centering
        \sublabel{(a)}{\includegraphics[width=0.7\textwidth]{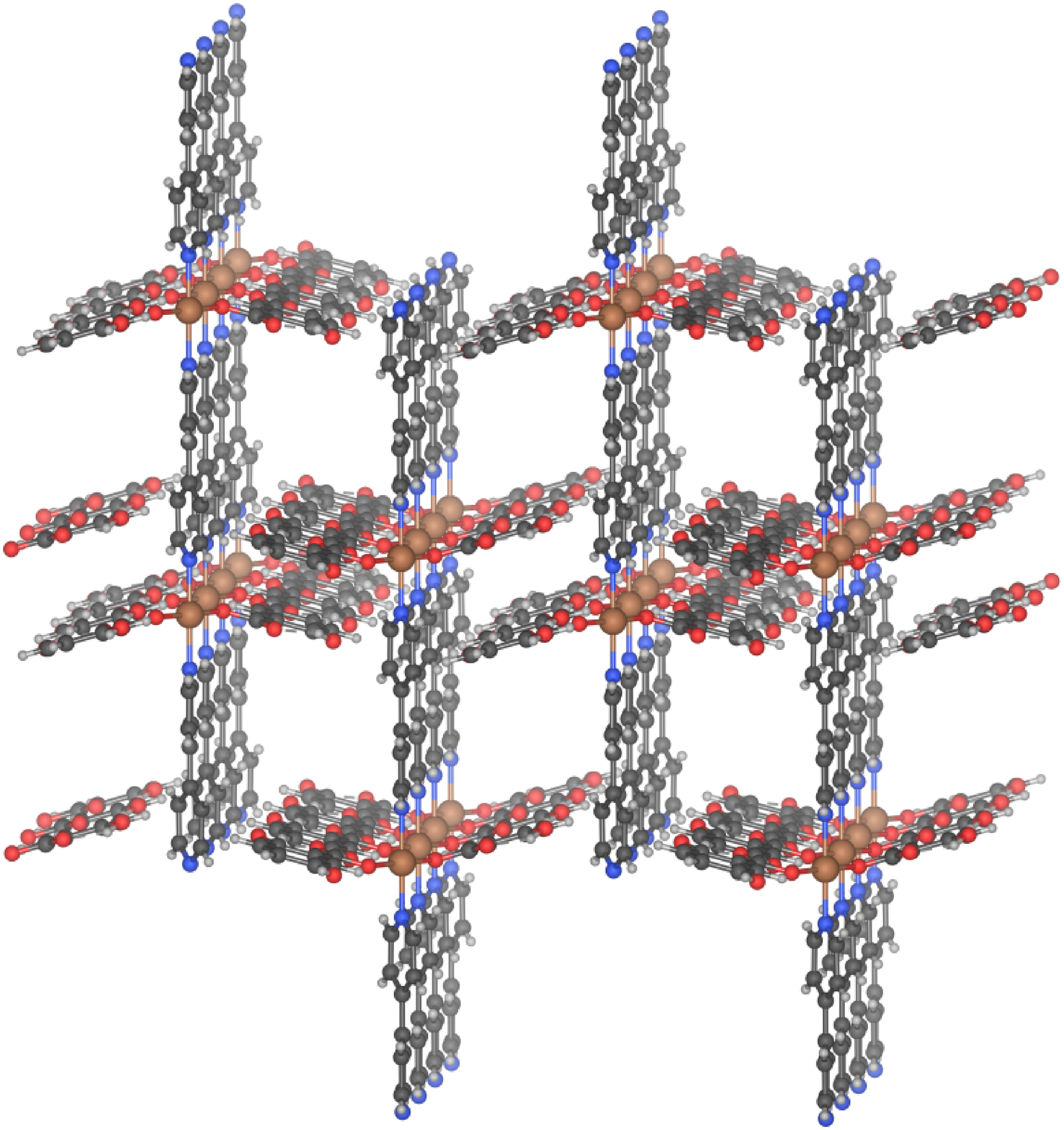}}
        \phantomsubcaption
        \label{fig:kitaura2003:structure}
    \end{subfigure}
    \begin{subfigure}[c]{.5\textwidth}
        %
        %
        %
        %
        %
        \input{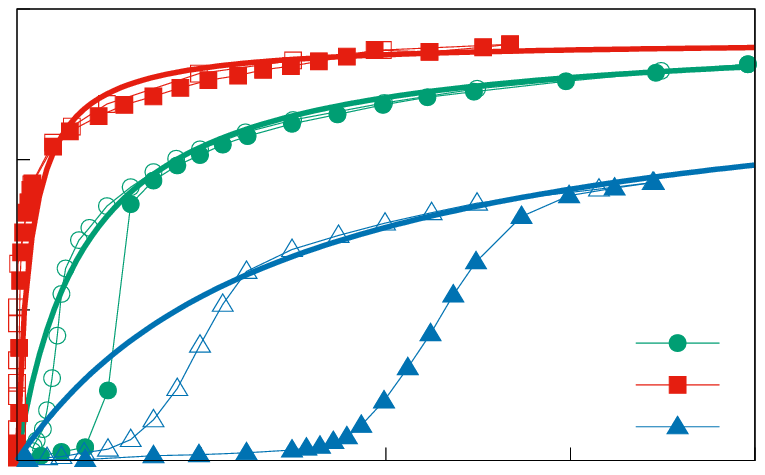}
        \label{fig:kitaura2003:isotherms}
    \end{subfigure}
    \caption{(\subref{fig:kitaura2003:structure}) \Cudhbc structure (from Ref.~\citenum{kitaura2003}).
    (\subref{fig:kitaura2003:isotherms}) Sorption isotherms and model isotherms
    fit at \SI{298}{K} in \Cudhbc for various gas compounds. Adsorption data are
    presented using filled symbols, and desorption data using empty symbols.
    Thick lines are Langmuir isotherms fitted at high loading. Experimental data
    published by \citeauthor{kitaura2003}\cite{kitaura2003}}
    \label{fig:kitaura2003}
\end{figure}

\begin{figure}[h]
    \centering
    \begin{subfigure}[T]{.45\textwidth}
        \sublabel{(a)}{\includegraphics[width=0.9\textwidth]{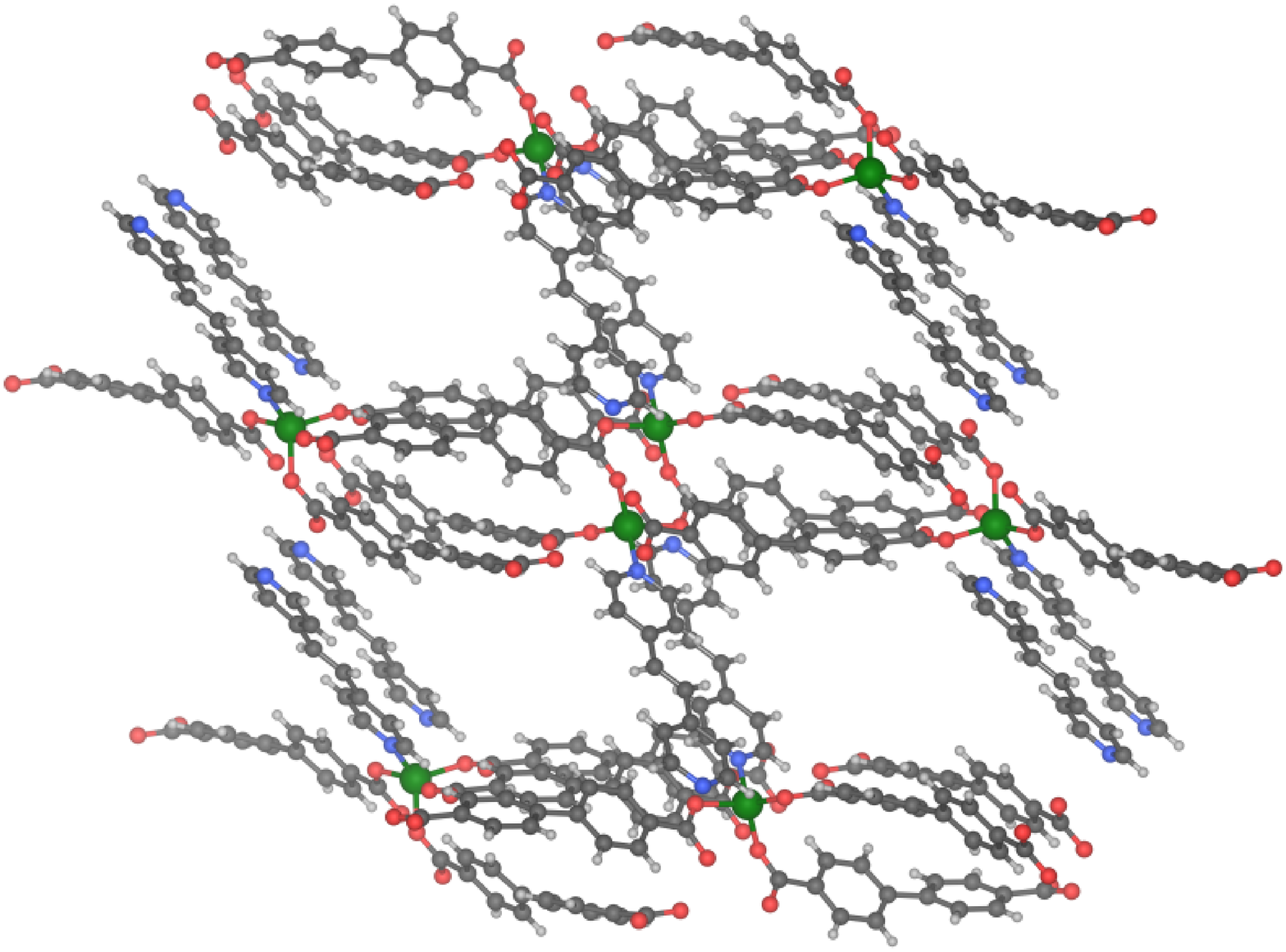}}
        \phantomsubcaption
        \label{fig:rpm3zn:structure}
    \end{subfigure}
    \begin{subfigure}[T]{.5\textwidth}
        %
        %
        %
        %
        \input{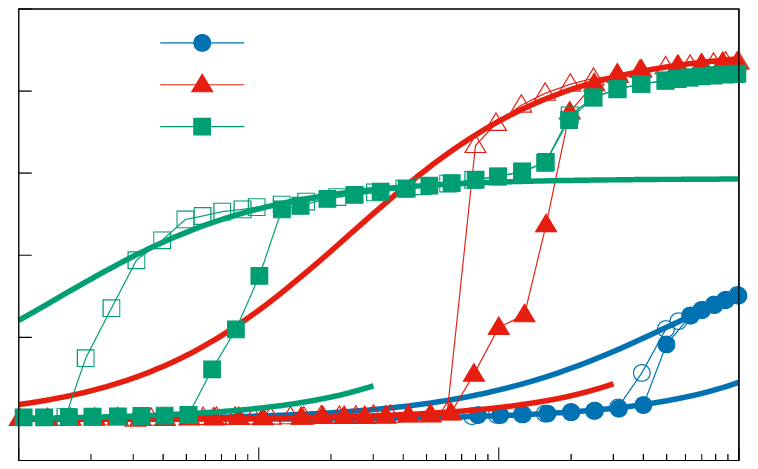}
        \phantomsubcaption
        \label{fig:rpm3zn:isotherms}
    \end{subfigure}
    \caption{(\subref{fig:rpm3zn:structure}) \RPMZn structure (from Ref.~\citenum{lan2009}).
    (\subref{fig:rpm3zn:isotherms}) Sorption isotherms at \SI{298}{K} for short
    alkanes in \RPMZn. Blue circles are for \ce{C2H6}, red triangles for
    \ce{C3H8}, and green squares for \ce{C4H10}. Filled symbols for adsorption,
    empty symbols for desorption. Thick lines are the open and closed phases fit
    of the isotherms. Experimental data published by
    \citeauthor{nijem2012}\cite{nijem2012}}
    \label{fig:rpm3zn}
\end{figure}

We present here two examples of co-adsorption of gas mixtures in metal--organic
frameworks with gate opening behavior, based on experimental data from the
published literature, comparing the predictions of IAST with those of OFAST. The
first example deals with the adsorption of $\ce{CO2}$, $\ce{CH4}$, and $\ce{O2}$
in the \Cudhbc MOF\cite{kitaura2003} (see figure~\ref{fig:kitaura2003}; dhbc =
2,5-dihydroxybenzoate; bpy = bipyridine). These isotherms correspond very
closely to the archetypal ``gate opening'' scenario described above. The
second example deals with linear alkanes (ethane, propane, and butane) adsorption in
\RPMZn MOF\cite{nijem2012}; Figure~\ref{fig:rpm3zn}
presents the framework structure of \RPMZn and relevant experimental adsorption
and desorption isotherms, from Ref.~\citenum{nijem2012}.

For both structures, we fitted the isotherms at high loading using a Langmuir
model for the isotherm in the open pores structure; and at low loading using a
Henry isotherm model for the closed pores structure. This choice is discussed in
the next section. The fit coefficients are given in supplementary information,
Tables~S1 and~S2. We performed OFAST
calculations using Wolfram Mathematica, the code is reproduced in the
supplementary information, and available as a full notebook online at
\url{https://github.com/fxcoudert/citable-data}. We computed the difference in
free energy between the two phases of the structures using these isotherms
models. We performed the pure IAST calculations using the PyIAST Python
package.\cite{simon2016} For the IAST calculations, we did not fit the isotherms
to a specific model, but rather the IAST equations were solved by numerical
integration and interpolation between experimental data points. At
partial pressures higher than the last point in the experimental
isotherm, that last point was used as saturation uptake. We only discuss
selectivity curves in the following section, as selectivity is often what people
are looking for when working with flexible porous media for the separation of
gas. The total and partial loading curves are also available in
supplementary information, Figures~S2
to~S7.

\subsection{Simple isotherms in \Cudhbc}

\Cudhbc is a textbook example of gate opening upon adsorption, with
single-component adsorption isotherms (reproduced in
Figure~\ref{fig:kitaura2003}) that clearly show the transition from a nonporous
(at low gas pressure) to a microporous (at higher pressure) host phase. From the
experimental data\cite{kitaura2003} we computed the free energy difference for
all the isotherms, and they all agree on the value of $-3.5 \pm
\SI{0.1}{kJ/mol}$. The exact values are given in the supplementary
Table~S3.

\begin{figure*}[ht]
    \input{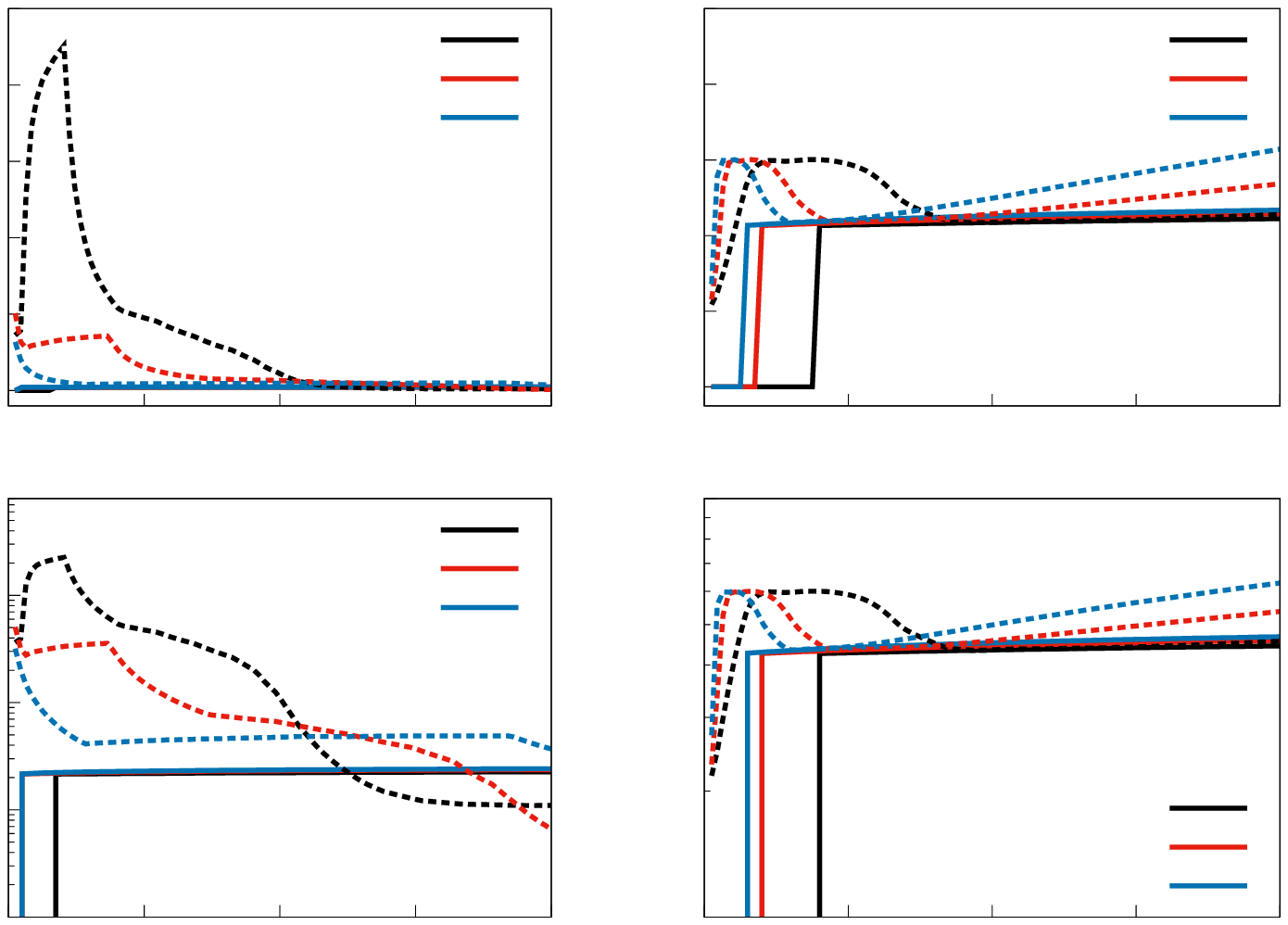}
    \caption{Comparison of IAST (dashed lines) and OFAST (plain lines)
    adsorption selectivity for \ce{CO2}/\ce{O2} (left) and \ce{CH4}/\ce{O2}
    (right) mixtures in \Cudhbc. The same curves are presented twice, using
    linear scale for the $y$ axis on the top panels, and logarithmic scale on
    the bottom panels.}
    \label{fig:cudhbc:iast-ofast}
\end{figure*}

Figure~\ref{fig:cudhbc:iast-ofast} presents the selectivity obtained with IAST
and OFAST for various gas mixtures and compositions in \Cudhbc. The adsorption
selectivity, calculated with OFAST, follow what one would expect: at low
pressure, the pores are closed and no gas enter the structure, making the
selectivity ill-defined --- the isotherms at low pressure cannot be fitted
and exploited for calculation of separation.
Then, at a pressure depending on the composition of the
gas phase, the gate opening transition occurs. At pressure higher than gate
opening pressure, the framework is in its open pore form, and the value of
selectivity depends on the relative saturation uptake of the two phases. The
selectivities observed are almost independent of the fluid mixture composition,
they are $\approx 20$ for \ce{CO2}/\ce{O2} and $\approx 4$ for \ce{CH4}/\ce{O2}
mixtures.

In stark contrast with this picture, the selectivities calculated by IAST are
clearly non-physical. All selectivity curves present a maximum in the pressure
range where gate opening occurs, with selectivities that can be several orders
of magnitude too high, with for example 2 000 instead of 20 for
\ce{CO2}/\ce{O2}. Even at higher pressure --- above the gate opening pressure
range --- the behavior is not identical to the OFAST calculations, because the
incorrect behavior at low pressure affects IAST directly in the integration of
the isotherms (Eq.~\ref{eq:iast2}). Moreover, the IAST selectivity for
\ce{CO2}/\ce{O2} presents a big jump around 40~atm when $y_{\ce{CO2}} = 0.1$.
Looking at the partial loading in figure~S4, we can
attribute this jump to an equilibrium displacement, \ce{O2} replacing \ce{CO2}
in the structure. This shows again the fact that IAST behaves as if the
structure was closed for \ce{O2}, while being open for \ce{CO2} at lower
pressure range. We thus confirm by a quantitative study the inapplicability of
IAST in flexible nanoporous materials.

\subsection{More complex isotherms: the case of \RPMZn}

\begin{figure*}[ht]
    \centering
    %
    %
    %
    \input{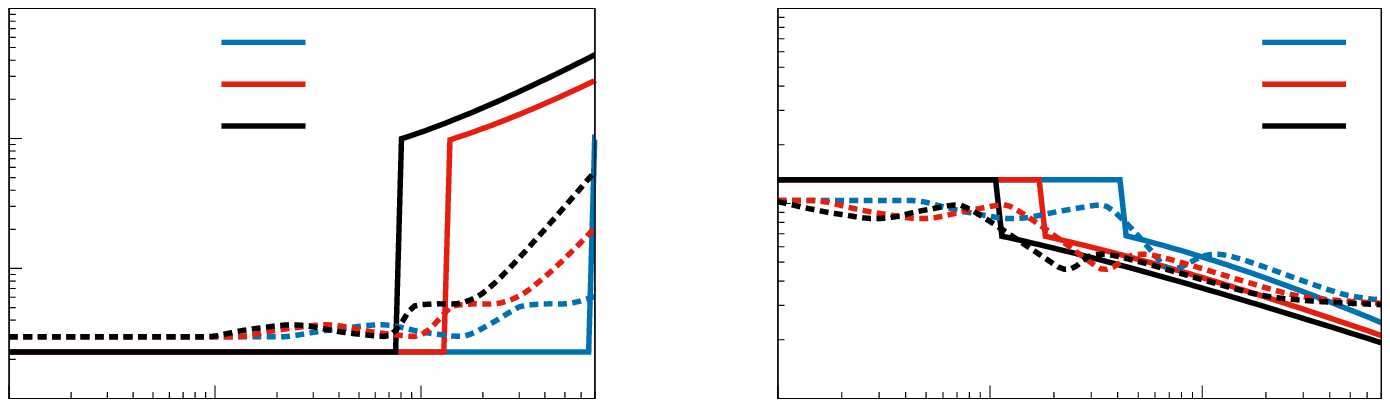}
    \caption{IAST (dashed lines) vs OFAST (plain lines) adsorption selectivity
    for \ce{C3H8}/\ce{C2H6} \emph{(left)} and \ce{C4H10}/\ce{C3H8}
    \emph{(right)} mixtures in \RPMZn at different compositions.}
    \label{fig:rpm3zn:iast-ofast}
\end{figure*}

We now turn to a second example of gate opening material, \RPMZn\cite{lan2009},
which presents more complex adsorption--desorption isotherms for short alkanes
(ethane, propane, butane) --- depicted on the right panel of
Figure~\ref{fig:rpm3zn:iast-ofast}. While adsorption of \ce{C2H6}, and \ce{C3H8}
in this material display a typical gate opening behavior, with a well-marked
single transition from a nonporous to a microporous phase, the adsorption of
\ce{C4H10} present two steps at \SI{0.01}{atm} and \SI{0.2}{atm}. There, the
first transition can be attributed to the structural transition (gate opening),
but the second one is of a different nature. Because there is no
hysteresis loop for the second step, and because it occurs for the larger and
more anisotropic guest molecule, it can be attributed to a fluid reorganization
(or fluid packing) transition inside the pores. Because experimental \emph{in
situ} characterization (such as single X-ray diffraction) would be necessary to
definitely affirm the character of this second step, we chose in the current
analysis to work in a reduced pressure range --- although the OFAST method
itself works with host materials with more than two phases. We thus fitted the
\ce{C4H10} isotherm using a Langmuir isotherm for pressures below \SI{0.2}{atm}.
The OFAST selectivity after this pressure will thus not be quantitatively
accurate, but will be sufficient for the physical insight we need. We also
performed tests by computing the selectivity under the assumption that the
second jump is due to fluid reorganization by using Langmuir-Freundlich
isotherms instead of single site Langmuir isotherm in the open phase, and the
selectivity only differs at pressures higher than \SI{0.2}{atm}.

From the \ce{C3H8} and \ce{C4H10} isotherms, we computed the free energy
difference between the nonporous and microporous phases, which we find to be
$\Delta F = -30.0 \pm \SI{0.1}{kJ/mol}$ (see supplementary
table~S4 for details). We did not use the \ce{C2H6}
isotherms for this purpose, as it has only limited data at high loading (at
pressure above 1 bar), which increases somewhat the uncertainty of the fit. We were
still able to fit the \ce{C2H6} isotherm with a Langmuir model and use it to
compute co-adsorption data, as the free energy difference of the two host phases
do not depend on the gas.

Figure~\ref{fig:rpm3zn:iast-ofast} displays the selectivity curves obtained with
IAST and OFAST for various gas mixtures and compositions in \RPMZn. Again, the
OFAST selectivity curve follows the expected behavior: it is constant at low
loading, where single-component isotherms follow the Henry model. In this
low-pressure region, adsorption is negligible and the selectivity cannot be
exploited in adsorption-based processes. However, we can see that because IAST
is using numerical integration, it is much more sensitive to details in the
single-component isotherms than the OFAST method, which is based on fits.

OFAST correctly describes the occurrence of gate opening, at a pressure which
depends on mixture composition but is in the range of the pure component gating
pressures. After gate opening, the selectivity jumps to its value in the open
pore framework. \ce{C3}/\ce{C2} mixtures have a behavior similar to that
observed in \Cudhbc, with a slowly growing (in log scale) selectivity at high
loading. On the other hand, OFAST selectivity for \ce{C4}/\ce{C3} mixture
displays a different behavior. The selectivity is lower after the transition
than before, and further decreases as the pressure and loading increases. This
is due to the fact that the single-component isotherms  in the open pore
structure cross, with \ce{C3H8} adsorbing more than \ce{C4H10} for pressure
bigger than \SI{0.03}{bar}. Thus, the low-pressure selectivity is reversed at
high pressure.

In contrast, the IAST fails to describe gate opening, with selectivity showing a
continuous evolution. Even the trends displayed by this evolution are in poor
agreement and make no physical sense, featuring non-monotonic evolution as a
function of pressure and composition. Even their high-pressure limit is often
far off from reality, as seen in the case of \ce{C3}/\ce{C2}.

\section{Conclusion}

Several published studies of fluid mixture
coadsorption in flexible nanoporous material use the Ideal Adsorbed Solution
Theory (IAST) method to predict the coadsorption behavior based on
single-component adsorption isotherms. This is an invalid application of IAST,
which is not adapted to flexible frameworks, as its very first hypothesis is
that the framework is inert during adsorption --- as clearly stated in the
derivation of the method in the seminal IAST paper.\cite{myers1965} However, the
IAST method can be adapted for frameworks presenting phase transitions induced
by adsorption by using the osmotic thermodynamic ensemble. This extension of
IAST to flexible materials is called Osmotic Framework Adsorbed Solution Theory
(OFAST).\cite{coudert2010} It allows the prediction of phases transitions upon
co-adsorption, as well as the details of the multi-component co-adsorption
isotherms, and is available in commercial software.\cite{vanassche2016} Moreover,
the use of OFAST with data at various temperatures allows one to produce
multi-dimensional {temperature, pressure, mixture composition} phase diagrams
for the flexible host.\cite{ortiz2012} Finally, while OFAST itself relies on the IAST
to describe adsorption in each phase of the host material, this method of
accounting for flexibility is not limited to IAST and can be used with other
adsorbed solution models, such as real adsorbed solution theory (RAST) or
vacancy solution theory (VST).

In this paper, we compared the results given by the IAST and the OFAST method
for co-adsorption of fluid mixtures in two different frameworks presenting a
gate-opening behavior. In both cases, the selectivities derived by the IAST
method are nonphysical and differ widely from the OFAST results, over- or
under-estimate the selectivity, sometimes by up to two orders of magnitude.
Moreover, we show that even without explicitly using IAST for calculations of
selectivity in flexible frameworks, one has to be cautious in comparing
single-component isotherms of different guests. Differences in step pressure of
stepped isotherms can lead to claims of strong selectivity using flexibility,
when applying --- without noticing it --- concepts that are valid only for
rigid host matrices.

\printbibliography

\end{document}